\documentclass[apl,twocolumn,showpacs]{revtex4}

\usepackage{graphicx}

\begin{document}

\title{
Microscopic Analysis of Low-Frequency Flux Noise in
YBa$_2$Cu$_3$O$_7$ Direct Current Superconducting Quantum
Interference Devices}

\author{D. Doenitz, R. Straub, R. Kleiner, D. Koelle}

\affiliation{ Universit\"at T\"ubingen, Physikalisches Institut --
Experimentalphysik II, D-72076 T\"ubingen, Germany }

\begin{abstract}
We use low-temperature scanning electron microscopy combined with
SQUID detection of magnetic flux to image vortices and to
investigate low-frequency flux noise in YBa$_2$Cu$_3$O$_7$ thin
film SQUIDs. The low-frequency flux noise shows a {\it nonlinear}
increase with magnetic cooling field up to 60 $\mu$T. This effect
is explained by the surface potential barrier at the SQUID hole.
By correlating flux noise data with the spatial distribution of
vortices, we obtain information on spatial fluctuations of
vortices on a microscopic scale, e.g. an average vortex hopping
length of approximately 10\,nm.
\end{abstract}

\pacs{68.37.Hk, 74.40.+k, 74.60.Ge, 74.76.Bz, 85.25.Dq}

\maketitle

The sensitivity of superconducting quantum interference devices
(SQUIDs) based on high transition temperature $T_c$
superconductors is in most cases limited by low-frequency $f$ flux
noise, with the spectral density of magnetic flux noise $S_\Phi$
typically scaling as $S_\Phi \propto 1/f$
\cite{ferrari94,koelle99}. The main source of this excess noise is
the thermally activated vortex motion in the high-$T_c$ thin films
forming the body of the SQUID or the superconducting input circuit
\cite{koelle99}. Low-frequency fluctuations induced by hopping of
vortices are strongly dependent on the nature and spatial
distribution of defects which act as pinning sites for vortices.
Therefore, integral measurements of flux noise and electric
transport properties give only very limited information on the
underlying vortex dynamics. Hence, techniques offering spatial
information are very useful to obtain a better understanding of
the basic mechanisms of low-frequency flux noise generation. \

Over the last two decades, low-temperature scanning electron
microscopy (LTSEM) has proven to be very successful in getting local
information on properties of superconducting thin films and
Josephson junctions, such as the spatial distribution of critical
temperature $T_c$ or critical current density $j_c$ and of Josephson
vortices in long junctions\cite{gross94}. Recently, this method was
successfully extended to the direct imaging of vortices in
YBa$_2$Cu$_3$O$_7$ (YBCO) SQUIDs\cite{koelle00}. Together with a
simultaneous measurement of the low-frequency flux noise of the
SQUIDs one obtains both images of the spatial distribution of
vortices and local information on noise properties of the devices
under investigation\cite{straub01}.

In this paper we investigate YBCO dc SQUIDs by LTSEM to study the
increase of magnetic flux noise with magnetic cooling field and to
analyze the mechanism of low-frequency flux noise generation on a
microscopic scale.

\begin{figure} [h] \noindent \center
\includegraphics[width=8cm]{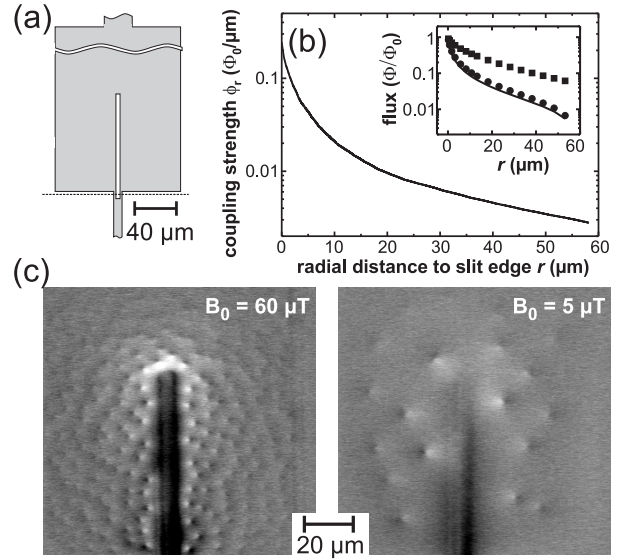}
\hfill \caption{\label{Abb} (a) SQUID washer design; dotted line
indicates grain boundary. (b) Coupling strength $\phi_r$ vs. radial
distance $r$ from the slit [numerical calculation for design in
(a)]. Inset: Magnetic flux $\Phi/\Phi_0(r)$ coupled by a single
vortex into the SQUID hole; numerical calculation for design in (a)
[squares] and for a circular washer [disks]; the line is obtained
from analytical calculation \cite{humphreys99} for a circular
washer. (c) $\delta \Phi$ images of the SQUID washer for two
different cooling fields $B_0$ ($T$=77\,K).}
\end{figure}

In our experiments, we use epitaxially grown YBCO thin film dc
SQUIDs of the so-called washer design [see Fig.~\ref{Abb}(a)]. The
$c$-axis oriented YBCO film has a thickness of 80 nm; the washer
size is $120 \times 305 \ \mu $m$^2$, with a 100 $\mu$m long and 4
$\mu$m wide slit. The 1 $\mu$m wide Josephson junctions are formed
by a 24$^\circ$ symmetric grain boundary in the underlying SrTiO$_3$
substrate. The YBCO SQUIDs are mounted on a magnetically shielded,
liquid nitrogen cooled cryostage of a SEM \cite{gerber97a} and read
out by a standard flux-locked loop (FLL) with 3.125 kHz bias current
reversal to eliminate $1/f$ noise due to fluctuations in the
critical current $I_c$ of the Josephson junctions.

For the spatially resolved measurements, the electron beam is used
as a local perturbation which induces an increase in temperature
$\delta T(x-x_0,y-y_0)$ on the sample surface (in the x-y-plane)
centered around the beam spot position ($x_0,y_0)$. The length scale
for the spatial decay of the thermal perturbation is set by the
beam-electron range $R\approx 0.5\,\mu$m for a typical beam voltage
$V_b$ = 10 kV\cite{gross94}. This gives a maximum increase in beam
induced temperature $\Delta T \approx$ 1 K at $(x_0,y_0)$ for a
typical beam current $I_b$ = 7\, nA. So-called $\delta\Phi(x_0,y_0)$
images are obtained by recording the e-beam induced flux change
$\delta \Phi$ in the SQUID as a function of the e-beam coordinates
$(x_0,y_0)$. To improve the signal to noise ratio, we use a
beam-blanking unit operating at typically 5\,kHz and the output
signal of the FLL, i.e., the e-beam induced flux change in the
SQUID, is lock-in detected. Additionally, the time trace or the
power spectrum of the FLL output signal can be recorded by a signal
analyzer.

The mechanism of imaging of vortices is explained in Ref.
\cite{koelle00} and can be briefly described as follows: The
e-beam-induced local increase in temperature induces a local
increase in the London penetration depth $\lambda_L$. Hence, the
screening currents circulating around a vortex are spatially
extended due to e-beam irradiation. If the e-beam is scanned
across a vortex this vortex is virtually dragged along with the
beam, i.e., it is displaced by some distance $\delta r$, if the
beam spot is within the radial distance $R$ from the vortex. This
displacement changes the amount of stray magnetic flux that a
vortex couples to the SQUID hole. Hence, scanning across a vortex
induces a negative (positive) flux change $\delta \Phi$ in the
SQUID if the vortex is moved away from (towards) the SQUID hole.
Figure \ref{Abb}(c) shows two examples of $\delta \Phi$ images
with vortices appearing as pairs of positive (bright) and negative
(dark) signals. These images were taken after cooling from above
$T_c$ to $T=$ 77 K in a static magnetic field $B_0$ = 60 $\mu$T
and 5 $\mu$T. The maximum displacement $\Delta r$ of a vortex is
of the order of the beam-induced change in $\lambda_L$ (typically
$\approx$ 20 nm in our experiments at 77 K)\cite{straub01}. The
displacement $\Delta r$ induces a signal $\delta \Phi =
\frac{\partial \Phi}{\partial r}(r) \times \Delta r$, with $r$
being the radial distance of the vortex from the SQUID hole.

For a given washer geometry, the {\it coupling strength}
$\phi_r(r)\equiv\frac{\partial \Phi}{\partial r}(r)$ is a function
of the vortex position only. Therefore, it can only be obtained from
a spatially resolved measurement. $\phi_r$ plays an important role
for the signal noise as well: The flux noise $S_{\Phi i}$ from a
moving vortex is $S_{\Phi i}= \phi_r^2(r_i) \times S_{ri}$, where
$S_{ri}$ is the spectral density of radial motion of the vortex at
distance $r_i$\cite{straub01}. \

By numerical simulation\cite{khapaev02} one can calculate the amount
of stray flux $\Phi$ that a single vortex couples into the SQUID
hole. For a given SQUID design, this value to a good approximation
depends for vortices to the left or right of the SQUID hole [see
Fig.\ref{Abb}(a)] only on the distance $r$ from the SQUID hole. The
dependence $\Phi(r)$ (normalized to the flux quantum $\Phi_0$) is
shown in the inset of Fig.~\ref{Abb}(b). We note that analytical
calculations of $\Phi(r)$ for circular washers\cite{humphreys99} are
in excellent agreement with our numerical simulation results [c.f.
inset of Fig.~\ref{Abb}(b)]. We find that $\Phi(r)$ strongly depends
on the washer geometry; e.g.~$\Phi(r)$ falls off with increasing $r$
much more rapidly for a circular washer as compared to a square
washer. Generally, $\Phi(r)$ quickly falls off from $\Phi_0(r=0)$
with increasing $r$ and approaches zero at the outer washer edge.
The derivative of this function, the coupling strength $\phi_r(r)$,
shows a similar behavior [see Fig.~\ref{Abb}(b)]. This has two
important effects on vortices located close to the SQUID hole: On
the one hand, the signal contrast in the image is high, since
$\delta \Phi\propto\phi_r$. On the other hand, the noise signal
produced by thermally activated motion of the vortex contributes
strongly to the total noise, since $S_\Phi\propto\phi_r^2$.\

\begin{figure} [h] \noindent \center
\includegraphics[width=7.5cm]{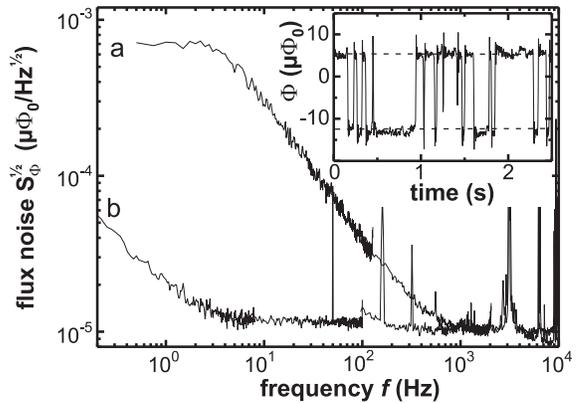}
\hfill \caption{\label{sp} Rms flux noise of YBCO dc SQUID at 77\,K
showing Lorentzian noise due to a single dominating fluctuator (a)
and low-frequency 1/$f$-noise (b). Inset: time trace of the SQUID
output with a single dominating fluctuator.}
\end{figure}

Vortices in superconducting thin films show thermally activated
hopping between two (or more) pinning sites. Due to the small
hopping length, the hopping process cannot be resolved directly by
LTSEM, but it appears as a {\it random telegraph signal} (RTS) in
the SQUID response due to a change in the magnetic flux coupled to
the SQUID hole. Sometimes, the SQUID signal is dominated by a single
fluctuator. In this case the RTS is clearly visible [see inset in
Fig.~\ref{sp}], producing a Lorentzian-shaped noise
spectrum\cite{machlup54} [see trace (a) in Fig.~\ref{sp}]. An
uncorrelated superposition of many RTSs leads to a 1/$f$-shaped
noise spectrum, provided the energies of the potential barriers
between the pinning sites are distributed uniformly\cite{dutta79}
[see trace (b) in Fig.~\ref{sp}].

\begin{figure} [t] \noindent \center
\includegraphics[width=7.5cm]{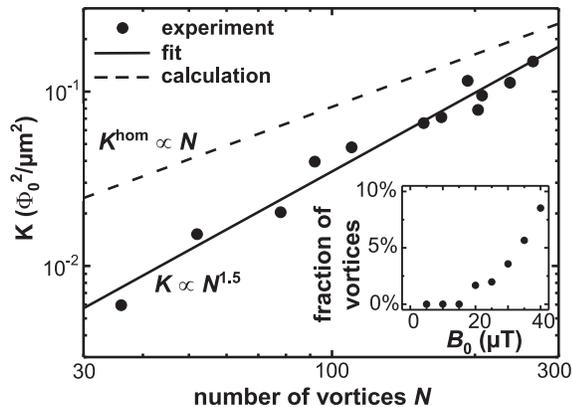}
\hfill \caption{\label{Rand} Sum of squared coupling strengths $K$
vs. number of vortices $N$. Dashed line: homogeneous vortex
distribution. Inset: fraction of vortices found at a distance $\le
6\,\mu$m from the slit edge for different cooling fields $B_0$.}
\end{figure}

For uncorrelated fluctuations of $N$ vortices trapped in the washer,
the total noise is
$$S_\Phi=\sum_{i=1}^N \phi_r^2(r_i)\times S_{ri}=\overline{S_r}\times K$$
where $\overline{S_r}$ is the average spectral displacement noise
power and $K\equiv\sum_i \phi_r^2(r_i)$ is the sum of the squared
coupling strengths. With the vortex positions $r_i$, obtained from
the $\delta \Phi$ images (for different $B_0$, i.e. different $N$)
and the calculated function $\phi_r(r)$ we determine $K(N)$ [see
Fig.~\ref{Rand}]. Surprisingly, we find $K \propto N^{1.5}$
instead of $K^{\rm hom} \propto N$, which is expected for a
homogeneous distribution of vortices, and which we calculate by
replacing the sum with the integral $N\times\int \phi_r^2(r) dr$.
Furthermore, we find $K^{\rm hom}>K$ for all $N$ [see dashed line
in Fig.~\ref{Rand}].

The reason for this deviation is a superproportional increase of the
density of vortices close to the SQUID hole with increasing $B_0$,
as imaged by LTSEM [see inset of Fig.~\ref{Rand}]. This result can
be explained by the formation of a surface barrier for vortex exit
at the washer edges: For low magnetic fields, i.e. few vortices in
the washer structure, the vortices are far apart from each other. In
this case the repulsive surface barrier potential\cite{brandt95} is
large compared to the repulsive vortex-vortex
interaction\cite{pearl64}, preventing formation of vortices close to
the SQUID hole during the cooling process. For higher fields, i.e.
increased vortex density, the vortex-vortex repulsion becomes
stronger, thus pushing vortices closer to the SQUID hole.

\begin{figure} [t] \noindent \center
\includegraphics[width=7.5cm]{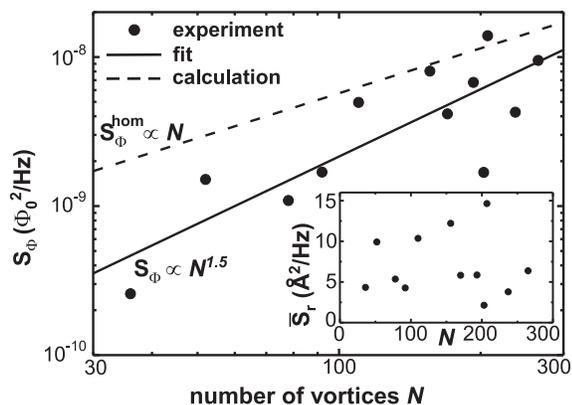}
\hfill \caption{\label{Sr} Spectral density of flux noise
$S_\Phi$(1Hz) vs. number of vortices $N$. Dashed line: homogeneous
vortex distribution. Inset: Average spectral density of radial
motion $\overline{S_r}$(1Hz) vs.~$N$, calculated from $K$ and
$S_\Phi$(1Hz).}
\end{figure}

With $S_\Phi \propto K$ and the scaling $K \propto N^{1.5}$ shown in
Fig.~\ref{Rand}, one expects $S_\Phi$ to increase stronger than
linear with the number of vortices $N$ or magnetic field. Figure
\ref{Sr} shows data for $S_\Phi$(1\,Hz) vs. $N$. Although
$S_\Phi(N)$ scatters quite strongly, it can be clearly seen that it
increases more like $\propto N^{1.5}$ than $\propto N$, in contrast
to previous measurements on larger SQUID washers\cite{ferrari94}. By
measuring $S_\Phi(1 $Hz) and determining $K$ from the $\delta \Phi$
pictures, one can calculate $\overline{S_r}(1$ Hz) using
$S_\Phi=\overline{S_r} \times K$ [see inset of Fig.~\ref{Sr}].
$\overline{S_r}(1$ Hz) is found to be 0.2-1.5 nm$^2$/Hz. For a
homogeneous vortex distribution, assuming $\overline{S_r}=0.7$\,
nm$^2$/Hz, one would get $S^{\rm hom}_\Phi \propto N$ with $S^{\rm
hom}_\Phi>S_\Phi$ in the investigated field range [see dashed line
in Fig.~\ref{Sr}].\

From these data it is possible to estimate the average vortex
hopping length $\overline{\Delta x}$ using a model based on the
following assumptions: (i) All vortices are situated in a symmetric
double well potential with fixed distance $\overline{\Delta x_r}$ of
local minima and variable barrier energy $E$. (ii) The barrier
energies $E$ are distributed uniformly, such that the highest
(lowest) characteristic frequency $f_c$\cite{dutta79} is 1000 Hz
(0.01 Hz). This condition provides an 1/$f$-shaped spectrum. Now
$\overline{\Delta x_r}$ can be calculated using the flux noise
theory developed in Refs. \cite{dutta79} and \cite{ferrari94}. The
result is $\overline{\Delta x_r} \approx$ (3 -- 8)\,nm. This
calculation gives the hopping length in {\it radial} direction.
Considering hopping in two dimensions, these numbers have to be
multiplied by the factor $\pi$/2 to obtain the overall average
hopping length $\overline{\Delta x} \approx$ (5 -- 13)\,nm. Although
this model uses some rather simple approximations it should produce
the correct order of magnitude of the vortex hopping length.

In conclusion, we combined vortex imaging by LTSEM with measurements
of low-frequency flux noise in YBCO dc SQUIDs to obtain information
on spatial fluctuations of vortices on a microscopic scale. We have
shown that a nonlinear increase of the flux noise with magnetic
cooling field can be explained by a surface barrier effect. Our
results clearly show that either strong pinning or avoding of
vortices trapped close to the SQUID hole is essential for reducing
low-frequency flux noise. Furthermore, our analysis yields an
average vortex hopping length around 10\,nm for our films. The
nature of defects producing pinning sites with hopping lengths on
this scale still has to be revealed.

We thank K. Barthel for fabricating the YBCO SQUIDs and John Clarke
for providing detailed layouts of the Berkeley SQUID readout
electronics. Furthermore, we gratefully acknowledge helpful
discussions with E.H. Brandt, T. Dahm, E. Goldobin, C. Iniotakis and
with R. Humphreys who also kindly provided unpublished results. This
work was supported by the Deutsche Forschungsgemeinschaft, the
Evangelisches Studienwerk e.V.~Villigst and the ESF program VORTEX.

\bibliography{Rauschen}
\end{document}